\pdfoutput=1

\documentclass[sigconf]{acmart}
\DeclareUnicodeCharacter{2212}{\textminus}

\acmConference[ ]{ }{ }{ }  
\acmYear{ }  
\setcopyright{none}  
\acmDOI{ }  
\acmISBN{ } 
\acmSubmissionID{}
\acmPrice{}

\AtBeginDocument{%
  }

\DeclareUnicodeCharacter{04D3}{\"a}

\copyrightyear{2025}
\acmYear{2025}
\setcopyright{rightsretained}
\acmConference[CHI EA '25]{Extended Abstracts of the CHI Conference on Human Factors in Computing Systems}{April 26-May 1, 2025}{Yokohama, Japan}
\acmBooktitle{Extended Abstracts of the CHI Conference on Human Factors in Computing Systems (CHI EA '25), April 26-May 1, 2025, Yokohama, Japan}\acmDOI{10.1145/3706599.3719834}
\acmISBN{979-8-4007-1395-8/2025/04}

\begin{document}

\title[Crafting Synthetic Realities]{Crafting Synthetic Realities: Examining Visual Realism and Misinformation Potential of Photorealistic AI-Generated Images}

\author{Qiyao Peng}
\orcid{0001-6964-4160}
\affiliation{%
  \institution{University of California, Santa Barbara}
  \city{Santa Barbara}
  \state{California}
  \country{USA}
}\email{qiyaopeng@ucsb.edu}

\author{Yingdan Lu}
\affiliation{%
  \institution{Northwestern University}
  \city{Evanston}
  \state{Illinois}
  \country{USA}}
\email{yingdan@northwestern.edu}

\author{Yilang Peng}
\affiliation{%
  \institution{University of Georgia}
  \city{Athens}
  \state{Georgia}
  \country{USA}}
\email{yilang.peng@uga.edu}

\author{Sijia Qian}
\affiliation{%
 \institution{University of California, Davis}
 \city{Davis}
 \state{California}
 \country{USA}}
 \email{sjqian@ucdavis.edu}

\author{Xinyi Liu}
\affiliation{%
  \institution{Northwestern University}
  \city{Evanston}
  \state{Illinois}
  \country{USA}}
\email{xinyiliu2028@u.northwestern.edu}

\author{Cuihua Shen}
\affiliation{%
  \institution{University of California, Davis}
  \city{Davis}
  \state{California}
  \country{USA}}
  \email{cuishen@ucdavis.edu}

\renewcommand{\shortauthors}{Peng et al.}

\begin{abstract}
 Advances in generative models have produced Artificial Intelligence-Generated Images (AIGIs) nearly indistinguishable from real photographs. Leveraging 30,824 AIGIs collected from Instagram and X (Twitter), this study combines quantitative and qualitative content analysis of a sample of 4,335 AIGIs to examine the photorealism of AIGIs through visual features related to perceived realism across content, human, aesthetic, and AI production features. We find photorealistic AIGIs often depict human figures, especially celebrities and politicians, with notable surrealism. Aesthetic professionalism is evident in staging and professional lighting. Only a small number of AIGIs show clear signs of generative AI, and AI production flaws were rare but varied. Our findings offer critical insights for understanding visual misinformation, mitigating potential risks of photorealistic AIGIs, and improving the responsible use of AIGIs.
\end{abstract}



\maketitle

\noindent
\section{Introduction}
\noindent The rise of AI-generated images (AIGIs) provides many opportunities for creative expression, but also raises concerns about misinformation risks through manipulated and synthetic visuals \cite{tc2023}. In 2023, AIGIs of Donald Trump’s arrest and Pope Francis wearing a Balenciaga puffer jacket circulated widely on Twitter, deceiving many users \cite{devlin_fake_2023}. On Reddit, an AIGI showing a blue plague in Russia, an event that has never happened, gained widespread attention \cite{designboom2023}. These incidents highlighted the risks and challenges of photorealistic AIGIs in misleading the public. 

AIGIs are produced entirely by artificial intelligence using algorithms that automate large-scale content production \cite{cao_comprehensive_2023}. Advances in generative models have enabled the creation of AIGIs that are nearly indistinguishable from real photographs with much lower cost and effort \cite{devlin_fake_2023}. Hausken \cite{hausken_photorealism_2024} has introduced the concept of AI photorealism, a style of images that emulate photography using advanced AI tools. Our study focuses on \textbf{photorealistic AIGIs}, and understanding their perceived realism through four feature-based dimensions: \textit{content features}, \textit{human features}, \textit{aesthetic features}, and \textit{AI production features}. Leveraging a large corpus of 30,824 AIGIs collected from Instagram and Twitter, we examined the visual features related to the perceived realism of a sample of 4,335 images. We developed a codebook with 42 categories to analyze AIGI features through content analysis. We also conducted a qualitative analysis on human-related surrealism and aesthetic professionalism. This paper provides essential insights for different stakeholders in AIGI production and consumption. It inspires future research on understanding the effects of AIGIs, and informs better AIGI practices and regulations by social media platforms, AI companies, and creative industries.

\section{Literature Review}
\subsection{AIGI and AI Photorealism}
Rooted in the tradition of photography that captures reality directly through the camera lens, photorealistic AI-generated images (AIGIs) redefine what a photograph looks like. Unlike smartphone photography which may use AI to enhance a real photograph, AIGIs are created entirely through artificial intelligence. Leveraging powerful generative models like Midjourney and DALL-E, photorealistic AIGIs mimic real photographs so convincingly that they are often indistinguishable from them \cite{devlin_fake_2023, lehmuskallio_photorealistic_2019, shen_this_2021}. This reflects the concept of photorealism — a visual style that simulates reality, which appears across various media such as computer graphics \cite{ferwerda_three_2003}, artistic works \cite{hausken_photorealism_2024}, and mixed and augmented reality \cite{pereira_photorealism_2021}. Photorealism involves a deliberate attempt to replicate the appearance of reality, challenging perceptions by blurring the line between the real and the artificial.

Photorealistic AIGIs complicate the definition of photorealism as these AI technologies far surpass previous methods in creating visuals with greater detail and may elicit more realistic judgment. Hausken \cite{hausken_photorealism_2024} conceptualized this as AI photorealism, a realistic visual style entirely reliant on AI tools. Unlike traditional photography or photorealism, AI photorealism introduces a fundamentally different approach to creating realistic images. However, a comprehensive and empirical-based understanding of AI photorealism in AIGIs is still lacking. Meanwhile, the proliferation of AIGIs, particularly in the form of highly realistic but false images, can accelerate the spread of misinformation. Thus, this paper looks into the specific features of photorealistic AIGIs and addresses their implications in the context of misinformation.

\subsection{Understanding Photorealistic AIGIs through Perceived Realism}
We explore photorealistic AIGIs by examining their visual features related to perceived realism. Pouliot and Cowen \cite{pouliot2007does} unpacked perceived realism into two dimensions. The first is factual realism, which pertains to the visual’s message style, whether it avoids distortion or artificial construction. The second is psychological realism, which refers to the viewer’s judgment of the semantic features of the messages and whether they appear plausible and could exist in the real world. With these two dimensions as a starting point, we further categorize the visual features related to the perceived realism of AIGIs into four dimensions: \textit{content features, human features, aesthetic features, and AI production features}.

Content features refer to the visual presentation of objects within the image, including their presence and combinations, such as text captions, humans, and animals \citep{peng_agenda_2023, qian_convergence_2024}. These features demonstrate how contextually accurate and believable the depiction of objects appears. Empirical evidence suggests that the number of unique objects, their natural appearance, and their combinations significantly influence perceptions of realism \citep{fan_automated_2014, fan_deeper_2020}. Building upon these studies, we focus on different types of objects and their combinations as our content features. 

Humans, as a specific category of objects, have significant implications for visual engagement and perceived realism \citep{li_is_2020, lu_mobilizing_2024, peng_what_2021}. Facial features, such as the presence of faces and identities, also influence social media engagement \citep{bakhshi_faces_2014, joo_visual_2014, munoz_image_2017}. In this study, we focus on the presence of human faces, their identities, and the interaction between humans and other elements to derive more comprehensive features of humans.

Visual aesthetics, which influence human perceptions of visual attributes \citep{peng_athec_2022, pieters_into_2010} and the diffusion of visuals \citep{bakhshi_red_2015, sharma_how_2024}, have been extensively explored in prior research using various indicators, such as colors, composition, and quality \citep{kanuri_standing_2024, lu_unpacking_2023, peng_athec_2022, peng_agenda_2023, qian_convergence_2024}. In the assessment of AIGIs and other visuals, color and quality are two crucial factors in considering realism \citep{chen_image-based_2024, cutzu_estimating_2003, lu_seeing_2023}. We build on these studies to capture the use of colors, perceived image quality, and aesthetic professionalism as aesthetic features.

AI production features reflect the development quality of the generative AI tool and can influence the audience’s ability to identify AIGIs. Flaws, such as irrational or illogical contents, contradictory elements, or content discontinuity, are often considered AI production features and are categorized under “rationality” or “artifacts” in previous studies of perceived realism \citep{chen_exploring_2023, groh2024human, lu_seeing_2023}. Compared with the traditional understanding of perceived realism, these production features are uniquely specific to the AI-generation context.

Grounded in theories related to these four dimensions, we propose the following research questions:

\textbf{\textit{RQ1:} What contents are present in photorealistic AIGIs?}

\textbf{\textit{RQ2:} What human features are present in photorealistic AIGIs?}

\textbf{\textit{RQ3}: What aesthetic attributes are present in photorealistic AIGIs?}

\textbf{\textit{RQ4:} What AI production features are present in photorealistic AIGIs?}

\section{Data and Methods}

\subsection{Social Media Data}

\begin{figure*}[htbp]
  \centering
  \includegraphics[width=\linewidth]{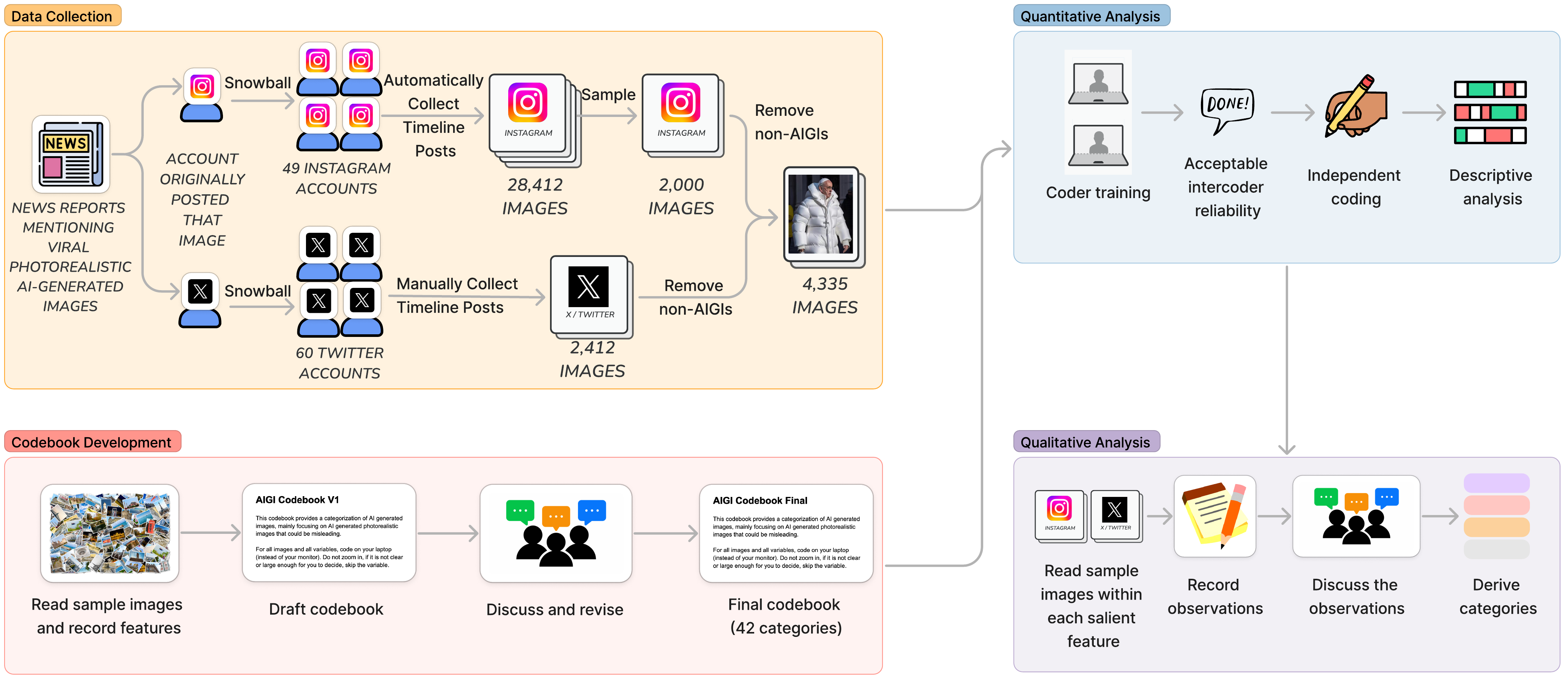}
  \caption{Data Collection and Analytical Methods}
  \Description{Flow Chart of Data Collection and Analytical Methods. The research design chart illustrates the process of data collection, codebook development, and analysis for a study on viral photorealistic AI-generated images. Data was collected from Instagram and Twitter accounts using snowball sampling, yielding 30,824 images. These images are sampled, filtered, and categorized through both quantitative and qualitative analyses. Quantitative analysis involves coder training, independent coding, and descriptive analysis. Qualitative analysis focuses on recording observations, discussing salient features of the images, and deriving categories.}
  \label{fig:design}
\end{figure*}

We collected AIGIs from Twitter and Instagram, as both platforms have substantial user bases where AIGIs are widely shared and engaged with and documented instances of viral AIGI misinformation \citep{chen2024twigma, devlin_fake_2023, petapixel2023}. We gathered data through a two-step process, as Figure~\ref{fig:design} shows. First, we identified AIGI accounts that frequently posted AI-generated images from July 2022 to August 2023. We started by examining news articles that mentioned viral photorealistic AIGIs to identify seed accounts on Twitter or Instagram where these AIGIs originated. Then, we snowballed and manually collected more AIGI accounts that were mentioned, forwarded to, or appeared on the recommendation lists of those seed accounts. We included AIGI accounts that were either self-identified as such or had at least 50\% of their most recent 20 images accompanied by captions, hashtags, or posts indicating that the images were created using specific generative tools. We terminated the sampling process once we consistently observed repeated mentions, forwards, and recommendations involving the same accounts. Ultimately, we included 49 Instagram accounts and 60 Twitter accounts.

From these accounts, we collected images from their timeline posts between July 12, 2022, and August 31, 2023. The start date was chosen to align with the public release of the most widely used AI image-generating tools, Midjourney and DALL-E. The end date was set for consistency, since our image collection began in September 2023, and we aimed to standardize the end time across all accounts. For Instagram, we used the 4K Stogram\footnote{4K Stogram is a desktop application for downloading photos, videos, and stories from Instagram using usernames, hashtags, or locations. \url{https://www.4kdownload.com/products/stogram-8}} tool to download the timeline posts and the associated images, accumulating a total of 28,412 images, from which we randomly selected 2,000 images for analysis. To match the Instagram sample, we manually downloaded approximately 60 images in chronological order from each Twitter account during the same time period as we did for Instagram, resulting in a total of 2,412 images (the combined total is 4,412: 2,000 from Instagram and 2,412 from Twitter).\footnote{Due to a change in the Twitter/X API, we were unable to easily automate the retrieval of a large number of images from this platform. As a result, we resorted to manually downloading them. In addition, during the image scraping process, two Instagram accounts and two Twitter accounts were excluded due to access issues. For some Twitter accounts that did not contain 60 images, we downloaded all available images instead.}

To minimize false positives and ensure the inclusion of as many AIGIs as possible, we manually reviewed 4,412 images in our dataset. We excluded 77 images that were most likely not AIGIs, including screenshots of social media or software interfaces, and images containing only text without any visual elements. The final dataset consists of 4,335 images for human annotation. While this approach may still include potential false positives, it represents the most effective strategy for building an AIGI dataset from social media at the early stages of AIGI development when such tools were newly available to the public.

\subsection{Quantitative and qualitative analysis}
We started our analysis with a quantitative content analysis. To develop a codebook for analyzing AIGI features related to perceived realism, we first drew 29 images from popular examples of misinformation-related AIGIs that went viral on social media, misled audiences, and attracted sizable media attention. These were collected via Google News searches for “AI photo,” “AI image,” and “AI fake” from December 2022 to July 2023, including controversial images such as the Pope in a Balenciaga puffer jacket \cite{Tolentino2023} and Donald Trump being arrested \cite{devlin_fake_2023}. Additionally, we randomly sampled 200 AIGIs from the original Instagram corpus (28,412 images total). One author specifically noted any features of these 229 images that could contribute to an AIGI’s potential to become misinformation. By combining well-known misinformation-related AIGIs with randomly sampled misinformation-prone AIGIs, we ensured that the codebook development was based on a diverse range of AIGIs.

Two additional authors independently skimmed the collected images based on the draft and provided feedback. Then, all six authors collaboratively refined and finalized the initial version of the codebook until no significant refinements were made. With this version, two trained coders independently analyzed a different random sample of 200 AIGIs drawn from the original Instagram corpus. Discrepancies in their coding were discussed and resolved until a satisfactory level of agreement was achieved. During the training phase, coders offered feedback, which the researchers used to further refine variable definitions and add clarifying examples. This iterative process enhanced the clarity and practical applicability of the final codebook. The coders achieved acceptable intercoder reliability across all variables in the final codebook (see Table~\ref{tab:codebook}).

The codebook starts by identifying whether an AIGI is photorealistic. After excluding non-photorealistic images, a total of 3,643 AIGIs (1,785 images from Instagram and 1,858 images from Twitter) were included for further analysis. For these photorealistic AIGIs, we coded 42 variables based on the four dimensions of perceived realism, 15 of which are first-order variables coded for each image, while the remaining 27 are second-order variables nested within selected first-order categories. All variables are binary variables. The major variables corresponding to the four theoretical dimensions of perceived realism in AIGIs are shown in Table~\ref{tab:codebook}. 

In the codebook, “Content Features" focuses on elements depicted in the images, such as textual captions, presence of humans, non-human living beings, and non-living objects. In addition, as AI-generated disaster images have been circulating on social media \citep{designboom2023}, potentially misleading the public and evoking fear or sympathy \cite{wang2021, gupta2013faking}, we included the portrayal of disasters and plagues. We also captured the surreal combinations that are unnatural pairings of situations, human, or objects not typically observed together in reality. Surreal combinations have deep roots in the art movement literature, where surrealism is characterized by juxtaposing unrelated, dream-like elements, people, or situations \cite{germain1974}. “Human Features" includes the presence of human faces and distinguishes identifiable individuals, particularly politicians and celebrities. “Aesthetic Features" encompass the use of colored images and perceived high image quality, based on the coder’s assessment of an image’s clarity and resolution. Finally, “AI Production Features" include the presence of any AI flaws or inconsistencies, and AI watermarks that explicitly indicate an image is AI-generated.

For AIGIs coded as containing AI production flaws, we followed Kamali et al. \cite{groh2024human} and manually identified three subcategories for more granular understanding. The first category is anatomical implausibilities, which refer to disproportionate human body features such as hands, eyes, and teeth that appear unnatural. The second category is functional implausibilities, which involve the misrepresentation of objects in the real world, often with distorted or unresolved details. The final category is violations of physics, which refers to inconsistencies with the laws of physics.

In addition to quantitative analysis, our content analysis and qualitative review of open notes from the human coding process revealed two prominent features across our collected AIGIs---human-related surrealism and aesthetic professionalism. To better understand and categorize these features, we evaluated a subset of images exhibiting each salient feature, using an open-coding strategy. After discussing these observations among all authors, we derived more granular subcategories within each feature.

\section{Results}
\subsection{Quantitative Analysis of AIGI Features}
From our quantitative analysis, we find that there are a large number of AIGIs (67.5\% on Twitter and 68.7\% on Instagram) contain human subjects. Surrealism combinations are also prevalent, at 58.2\% on Twitter and 60.8\% on Instagram, reflecting a substantial incorporation of imaginative and unreal combinations in the visuals. Chi-square tests confirm similar patterns for content features across platforms, with minor variations, as detailed in Appendix~\ref{sec:Appendixc}. Among images containing humans, 73.4\% of Instagram images and 71.0\% of Twitter images showed that human faces (\textit{$\chi^2$}(1) = 0.54, \textit{p} > 0.05). Among these, more than 10\% are identifiable as politicians or celebrities, with significant platform differences (\textit{$\chi^2$}(1) = 4.73, \textit{p} < 0.05). Instagram features more politicians, such as Donald Trump, Barack Obama, and Queen Elizabeth II, while Twitter displays more celebrities, such as Elon Musk, Mark Zuckerberg, and Kanye West. 

Photorealistic AIGIs also exhibit high levels of aesthetic features on both Instagram and Twitter. 98.0\% of Twitter photorealistic AIGIs and 93.0\% of Instagram photorealistic AIGIs are colored images, and over 92\% of photorealistic AIGIs on both platforms display high perceived image quality.

Less than 9\% of photorealistic AIGIs on both platforms contain observable AI production flaws. Among the 131 flawed AIGIs identified on Instagram and 150 on Twitter, approximately 84\% and 83\% of them exhibit anatomical implausibilities, such as misaligned eyes, unnatural arrangements or numbers of fingers, and twisted limbs bending at impossible angles. Beyond this, AIGIs in the Instagram sample show significantly higher proportions of functional implausibilities (e.g., distorted or illogical object interactions, Instagram 63.4\%, vs. Twitter 24.0\%), and violations of physics (e.g., misaligned shadows and depth errors, Instagram 26.7\%, vs. Twitter 13.3\%) compared to those of Twitter, highlighting platform-specific variations in the prevalence and nature of AI production flaws. In addition, only 1.5\% of Instagram AIGIs contain AI watermarks that signal AI production, while this figure is lower on Twitter, at 0.8\%.

\subsection{Human-related Surrealism}
Through our qualitative analyses of human-related surrealism in photorealistic AIGIs, we find that AIGIs depict humans in surrealistic settings, including surreal: 1) physical characteristics, 2) behaviors and interactions, and 3) contexts.
\begin{figure}[h]
  \centering
  \includegraphics[width=0.9\linewidth]{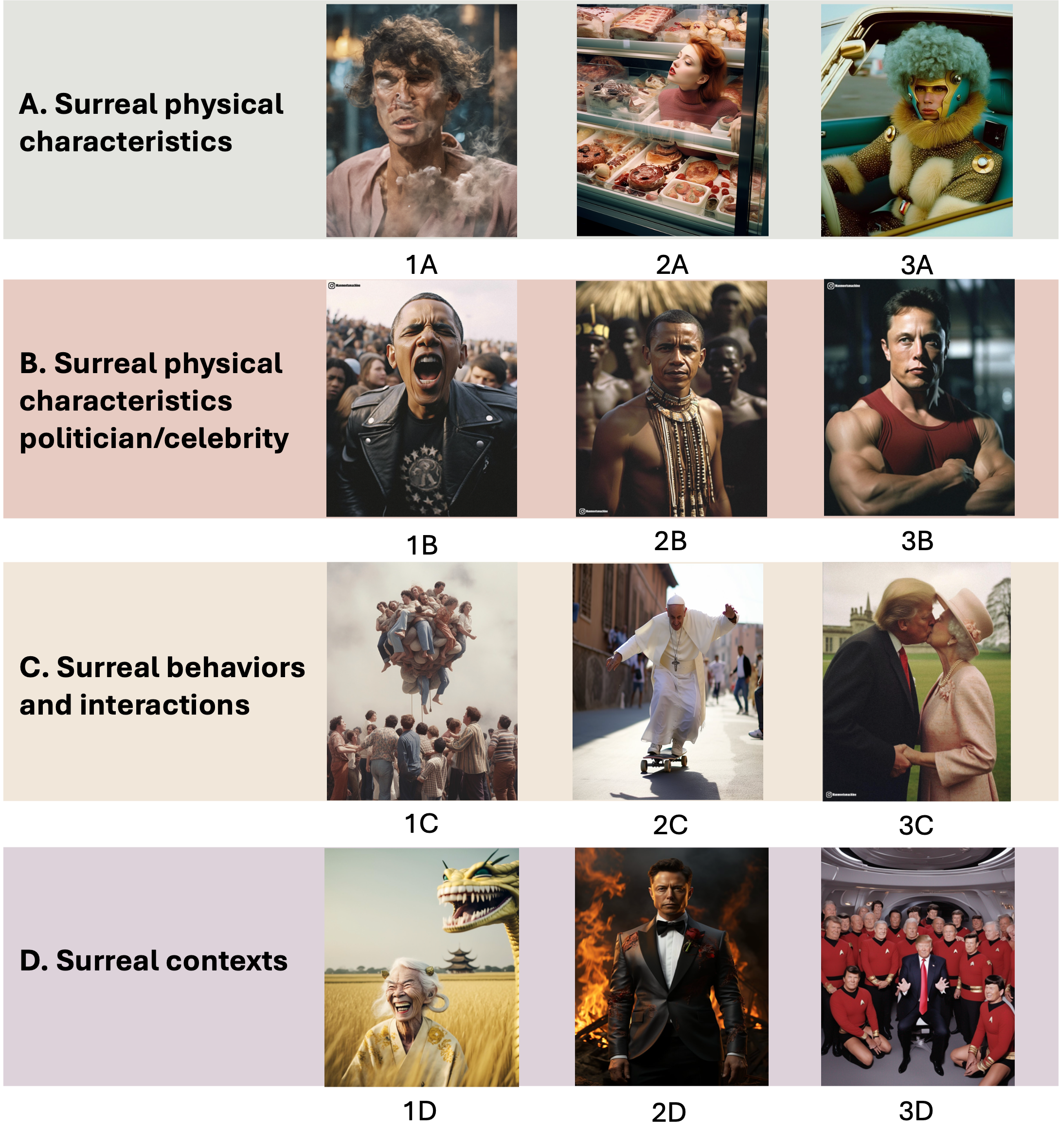}
  \caption{AIGI examples of surrealism.}
  \Description{This figure shows AIGI examples of human-based surrealism. The image presents different types of human-related surrealism: A depicts surreal physical characteristics of ordinary individuals; B features surreal physical characteristics of politicians and celebrities; C shows surreal behaviors and interactions; and D portrays surreal contexts. Each category contains three representative images with exaggerated or surreal elements, highlighting the surrealism-based manipulation of appearances, actions, and environments.}
  \label{fig:surrealism}
\end{figure}

\subsubsection{Surreal Physical Characteristics}
We find a large body of human-based surreal AIGIs portraying a single person with surreal physical characteristics, including facial abnormalities, body abnormalities, and clothing incongruence. For facial abnormalities, AIGIs usually present a distorted or asymmetric face, or an exaggerated expression with unlikely intensity (Panel 1A in Figure~\ref{fig:surrealism}). For body abnormalities, AIGIs usually present incomplete bodies or over-engineered functional body features (Panel 2A). AIGIs may also present clothing or costumes that are normally not seen in real life, including unusual clothing segments and accessories, or not commonly expected (Panel 3A).

Politicians are highly likely to have their physical characteristics with surreal representations in AIGIs. Our dataset reveals a salient presence of the former U.S. Presidents such as Donald J. Trump and Barack Obama, whose physical features were surrealistically altered. One such image, shown in Panel 1B in Figure~\ref{fig:surrealism}, portrays Barack Obama in a leather jacket at a public protest or concert, where his facial expression of excitement is highly exaggerated. This resonates with prior research suggesting that images of politicians in non-political settings and displaying emotions may elicit more online engagement \citep{peng_what_2021}. Another portrayal of Obama in Panel 2B, features him wearing indigenous attire, including necklaces and a chest ornament, which deviates from the traditional attire typically worn by Obama.

In addition to politicians, celebrities from various fields, such as Elon Musk, a globally renowned entrepreneur, are prominently featured in these AIGIs. As shown in Panel 3B, the excessively large muscles and the absence of natural body textures exemplify the surrealism in Musk’s visual representation. Combining this formidable physique with his serious facial expression, the image portrays a heightened sense of masculinity and a strong-man image of Musk.

\subsubsection{Surreal Behaviors and Interactions}

Human-related surrealism is also reflected in the portrayal of human behaviors and interactions, which portrays either supernatural behaviors or normal behaviors in unrealistic or even bizarre occasions. For example, Figure~\ref{fig:surrealism} Panel 1C shows a group of people clumping together and floating in mid-air, which violates the law of physics. Additionally, surreal behaviors emerge through the juxtaposition of a powerful figure and an unexpected action. For example, Panel 2C shows Pope Francis engaging in skateboarding, a highly unlikely activity given his physical condition and role as a religious leader. Surrealism also occurs when depicting multiple politicians or celebrities in their interactions. Panel 3C shows Donald Trump and Queen Elizabeth II kissing, which is both implausible and absurd. We found similar intimate behaviors depicted in AIGIs that involve other politicians and celebrities.

\subsubsection{Surreal Contexts}

Finally, surrealism occurs when humans are placed in nearly impossible or supernatural contexts. As Panel 1D shows, an elderly woman in a traditional outfit with a joyful expression is accompanied by a savage-looking dragon. In addition to ordinary people, we also find AIGIs that place politicians and celebrities in surreal contexts. For example, Panel 2D shows Elon Musk, in a formal but burnt suit, standing confidently in a close distance with flames and destruction. The mixture of elements that represent order and elegance (suit, rose, calm expression) with those that represent chaos (flame, destruction, burnt clothes) exemplifies surrealism and may portray a masculine and authoritative image of Musk. Groups of individuals can also serve as surreal contexts. In Panel 3D, Donald Trump is surrounded by people dressed in Star Trek-style uniforms, with incongruent head sizes, positions, and some in bizarre outfits. This depiction of surrounding individuals seems more surreal in the spaceship-like background. 

\subsection{Aesthetic Professionalism}
\begin{figure}[h]
  \centering  \includegraphics[width=0.9\linewidth]{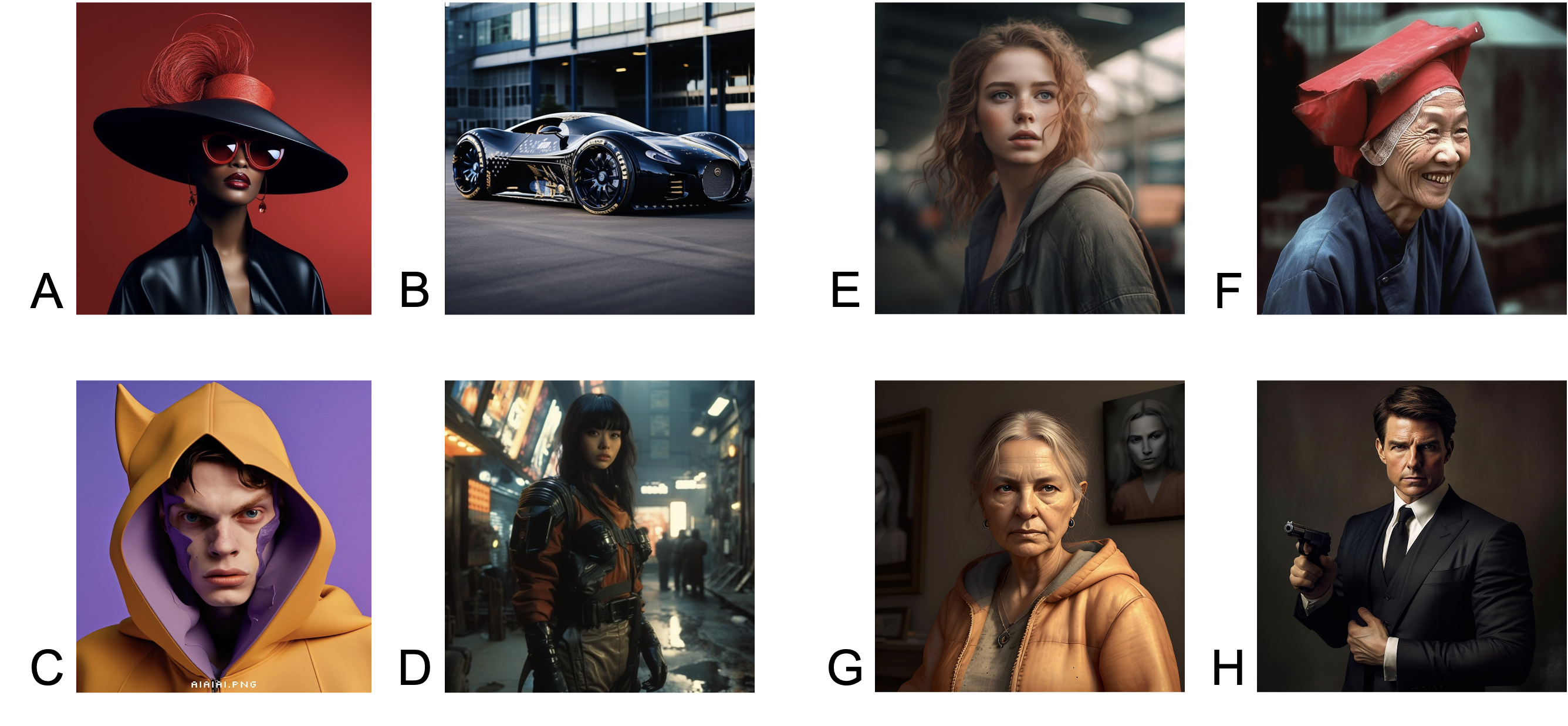}
  \caption{Staging and professionalism examples.}
  \Description{The image displays eight AI-generated visuals with a high level of aesthetic professionalism, constructed by staging and professional-quality lighting. The images are labeled A through H, demonstrating a range of staging and professionalism. These include a fashionable figure in a wide-brimmed hat (A), a futuristic sports car (B), a stylized portrait of a person in a hoodie (C), a cyberpunk character (D), a woman in an industrial setting (E), an older woman smiling in a red headscarf (F), an older woman indoors in a contemplative pose (G), and a man in a suit holding a gun, resembling a spy or action figure (H).}
  \label{fig:prof}
\end{figure}

Our collected AIGIs also reflect a high level of aesthetic professionalism, constructed by the staging and professional-quality lighting. Staging, or staged photography, poses the subject with controlled theatrical effects, lighting, and settings in photography \citep{castro_diaz_visual_2019, neumuller_routledge_2018}. As Panel A in Figure~\ref{fig:prof} shows, the model was arranged in a deliberate position, with a fashionable or exaggerated costume, a pure-color background, and crafted studio lighting. Non-human objects can also be the subject of a staged photograph. Panel B represents a staged sports car, positioned with artificial lighting and at a particular angle. The car’s shiny exterior, crafted paint, and sleek finishes make it luxurious and visually appealing.  

The staging feature corresponds to how AIGIs generally follow conventional photographic standards of professional quality. For example, Panels C and D tend to have high resolution and vivid detail. Notably, many of the AIGIs have one focal subject and a shallow depth of field, as shown in Panels E and F. This refers to a photographic technique where only a small portion of the image is in sharp focus, while the rest appears blurred. The primary subject, usually in the foreground, remains sharp and clear, while the background and sometimes even the foreground elements become out of focus. This technique is often used to draw attention to the main subject and minimize distractions. A large percentage of AIGIs follow this convention, demonstrating signs of professional quality.

Another indicator of aesthetic professionalism in photorealistic AIGIs is the expert use of lighting. For instance, in Figure~\ref{fig:prof} Panels G and H, the light is well arranged, often lit from one side to create a dramatic effect of light and shadow in portraits. One image may use "Rembrandt lighting" to create a small, triangular patch of light on the cheek of the subject, while half of the face is lit and the other half is in shadow. These dramatic lighting effects are characteristic of professional portrait photography, suggesting that these AIGIs adhere to high standards of visual aesthetics.

\section{Discussion}
As one of the first thorough examinations of photorealistic AIGIs circulating on different social media platforms, this study empirically investigates the visual features contributing to perceived realism and the potential implications of these realistic depictions. Grounded in the concepts of photorealism and two dimensions of perceived realism, we conceptualize AIGI photorealism into four dimensions. Through a comprehensive analysis of 4,335 AIGIs from Twitter and Instagram using quantitative content analysis and qualitative image analysis, our descriptive findings show that photorealistic AIGIs often depict human figures, especially celebrities and politicians, with a high degree of surrealism. The aesthetic professionalism is evident in the staging and professional lighting. Only a small proportion show overt signs of generative AI, and AI production flaws were rare but diverse, including anatomical, functional, and physical flaws. 

Collectively, the features of photorealistic AIGIs uncovered in our analysis point to their very high potential to become misinformation --- either intentionally through disinformation campaigns or inadvertently as they spread organically via online platforms, out of their intended context. First, the prevalence of humans, especially politicians and celebrities, coupled with a highly professional aesthetic style, provides the perfect recipe for mimicking photorealistic images usually found in news media. For instance, the surreal presentation of political figures may negatively influence the public’s perceptions and trust in political figures by "kicking up'' \citep{wagg_because_1998}. Second, the AIGIs we analyzed all lack discernible AI watermarks, therefore making it difficult for online users to trace where these images are from, especially when such images would be placed in a misleading context or blend with a high level of surrealism. Third, previous research has consistently shown that the presence of human faces \cite{li_is_2020, lu_mobilizing_2024} and high production quality \cite{li_is_2020}, which we found in AIGIs, contribute to higher user engagement such as likes and retweets. In other words, the photorealistic AIGIs we analyzed not only look real and resemble the images in credible news reports in both content and style, but they are also likely to engage online users and potentially reach a large audience. 

Our findings thus greatly complement and inform existing research that focuses on the  “implausibilities” of AIGIs for identification and user literacy interventions \cite{groh2024human}. Current approaches to distinguishing AIGIs from authentic photos often focus on the “implausible” parts such as incorrect human anatomies, which occur infrequently, are difficult to spot, and are likely to become even rarer as Generative AI technologies continue to evolve. Instead, our approach examines the opposite – the AIGI elements that occur frequently and make AIGIs appear real to human perceptions. Such perceptions of realism expose areas of user vulnerability to synthetic visuals, so platforms can design targeted user interventions and quarantine suspicious high-realism visual misinformation before they spread widely. 

Taken together, we argue that the direct and indirect misinformation potential of photorealistic AIGIs should not be underestimated, and it is imperative to build resilience against it. Building on this observational study, more research can leverage user experiments to systematically test how human, content, aesthetic and AI production features influence users' perceptions of realism in misinformation AIGIs. Future research can also examine other privacy and security outcomes, including trust, sharing intentions, and changes in attitudes or behaviors. In addition, our findings may inspire future research to integrate textual analysis of captions and embedded text with visual content analysis, as well as considering other contextual social media features, to provide a more holistic understanding of how AIGIs are presented and interpreted in a misinformation context. By pursuing these research directions, future studies can provide a more comprehensive understanding of AIGIs and their impact on viewer perceptions, ultimately contributing to more effective management and mitigation of misinformation in various media environments.

Finally, we emphasize the following design recommendations to improve the use and regulation of photorealistic AIGIs. We argue that addressing AI-generated visual misinformation should not be the sole responsibility of individual users, but rather a shared responsibility at a broader level, involving AI developers, social media companies, and other key stakeholders. A potential solution is the implementation of mandatory labels or provenance cues, which are reliable and verifiable information about the origin and evolution of digital content \cite{gregory2022deepfakes}. One example of provenance cues is the Coalition for Content Provenance and Authenticity (C2PA)\footnote{See: \url{https://c2pa.org/}} led by Adobe and others, allowing consumers to access the metadata of each image, trace the content’s origin and modifications, and verify authenticity through verification badges, assuring them of the content’s authenticity. These cues can be integrated within the algorithm itself, which ensures that the labels remain intact across different platforms. Such markers would provide an immediate visual cue to users, indicating that the content they are viewing was generated by AI, thereby increasing discernment and reducing the risk of misinformation. By clarifying the origins of digital content, these platforms can cultivate a more informed and critically engaged user base.

In addition, our study highlights the critical need to closely monitor and regulate AI-generated content (AIGCs) that involves human subjects, particularly when it includes identifiable faces of public figures such as politicians and celebrities. AI and social media companies should implement strict rules that govern the inclusion of real person's images in AI-generated content, ensuring that such usage does not contribute to the dissemination of false information or harm the reputations of those depicted.

\begin{acks}
This work was partially supported by the National Science Foundation (CNS-2150716, CNS-2150723). We thank Ziqi Liu and Yundi Zhang for their superb research assistance. We also thank members of the Computational Multi-Modal Communication (CMMC) Lab, Computational Media and Politics (COMAP) Lab, and participants of the 74th Annual Conference of the International Communication Association Images-as-Data panel for their valuable feedback. 
\end{acks}

\bibliographystyle{ACM-Reference-Format}
\bibliography{AIGIB1}

\clearpage
\appendix
\onecolumn
\section{Definitions of Photography, Traditional Photorealism and AI Photorealism}\label{sec:Appendixa}
\begin{table}[!ht]
    \centering
    \begin{tabular}{p{4cm}p{11cm}}
    \toprule
         & \textbf{Definition} \\
    \midrule
         \textbf{Photography} & The art or practice of using a camera to capture real scenes and moments, relying on the physical world, context, and timing. \\
         \textbf{Traditional Photorealism} & A genre of art that strives to create images so realistic they mimic actual photographs. \\
         \textbf{AI Photorealism} & The development of photorealism in the age of AI --- the style of images that emulate photography using advanced AI tools, with images created entirely by AI algorithms.\\
         
    \bottomrule
    
    \end{tabular}
    \caption{Definitions of Photography, Traditional Photorealism, and AI Photorealism}
    \Description{Definitions of Photography, Traditional Photorealism, and AI Photorealism}
    \label{tab:definition}
\end{table}
\twocolumn

\onecolumn
\section{Codebook and Intercoder Reliability}\label{sec:Appendixb}

\begin{table*}[!ht]
    \centering
    \resizebox{\textwidth}{!}{
    \begin{tabular}{p{4cm}p{11cm}p{1cm}}
    \toprule
         \textbf{Variables} & \textbf{Definition} & \textbf{Cohen’s Kappa} \\
         \midrule
        \textbf{Content features}\\
    \midrule
        1. Human & The presence of an object that is human being & .91\\
         2. Human-like & The presence of an object that has the shape of human-beings but not human or hard to be categorized as human, e.g., ghosts, aliens, human-shape robots etc. & .68\\
        3. Non-human living beings  & The presence of an object that has life but is not a human or human-like object. & .84\\ 
       4. Non-living objects & The presence of an object that has no life. & .96*\\
        5. Add-on captions & The presence of a caption with understandable content that deliver a specific message, yet are an addition to the image and do not inherently belong to the image itself. & .97*\\
        6. Fictional captions & The presence of a caption composed of fictional characters or letters, representing languages that do not exist in the real world, thus making their meanings unidentifiable. & .83\\
        7. Other captions  & The presence of a non-distorted in-image caption. & .68\\
        8. Disaster  & The presence of an event or situation that causes significant and widespread damage. & 1.00*\\
       9. Surrealism combination & The  presence of unnatural pairings of situations, human, or objects not typically observed together in reality. & .79\\
    \midrule  
       \textbf{Human features}\\
    \midrule
       1. Face & More than 50\% of the face is identifiable and is clear and large enough for viewers to detect its identity or emotions.  & .96\\
       1.1 Identity & Human identity is identifiable based on the presented human face.  & 1.00*\\
       1.1.1 Politician  & The presence of any politicians that can be recognized based on the best of knowledge. Politicians refer to people who work in government, making rules and decisions that affect the country or region they represent. & 1.00*\\ 
       1.1.2 Celebrity & The presence of any celebrities who are not political figures and can be recognized based on the best of knowledge. & 1.00*\\
        \midrule 
\textbf{Aesthetic features}\\
    \midrule    
        1. Colored images  & If it is a colored image. & .85\\
        2. High perceived image quality & The image is clear, with high resolution. & .97*\\
    \midrule
    \textbf{AI Production features}\\
        \midrule
     1. AI production flaw & The presence of any obvious production flaws can be told from unequivocal evidence that this image is made by AI. & .76\\
        2. AI watermark & The presence of a watermark that explicitly indicate the image is AI-generated. & .66\\
    \bottomrule
    \multicolumn{3}{p{\textwidth}}{\textit{Note: * indicates that we used percentage agreement instead of Cohen’s Kappa because the skewed distribution of our 0/1 coding can make Cohen’s Kappa less reliable. When one category is infrequent, kappa values can be misleadingly high or low.}} \\

    \end{tabular}}
    \caption{Codebook of Perceived Realism Variables for Photorealist AIGIs}
    \Description{Variables, Definitions, and Cohen's Kappa of Each Variable.}
    \label{tab:codebook}
\end{table*}
\twocolumn

\onecolumn
\section{The Percentage of AIGI Visual Features}\label{sec:Appendixc}
\begin{figure*}[!ht]
  \raggedright  \includegraphics[width=\linewidth]{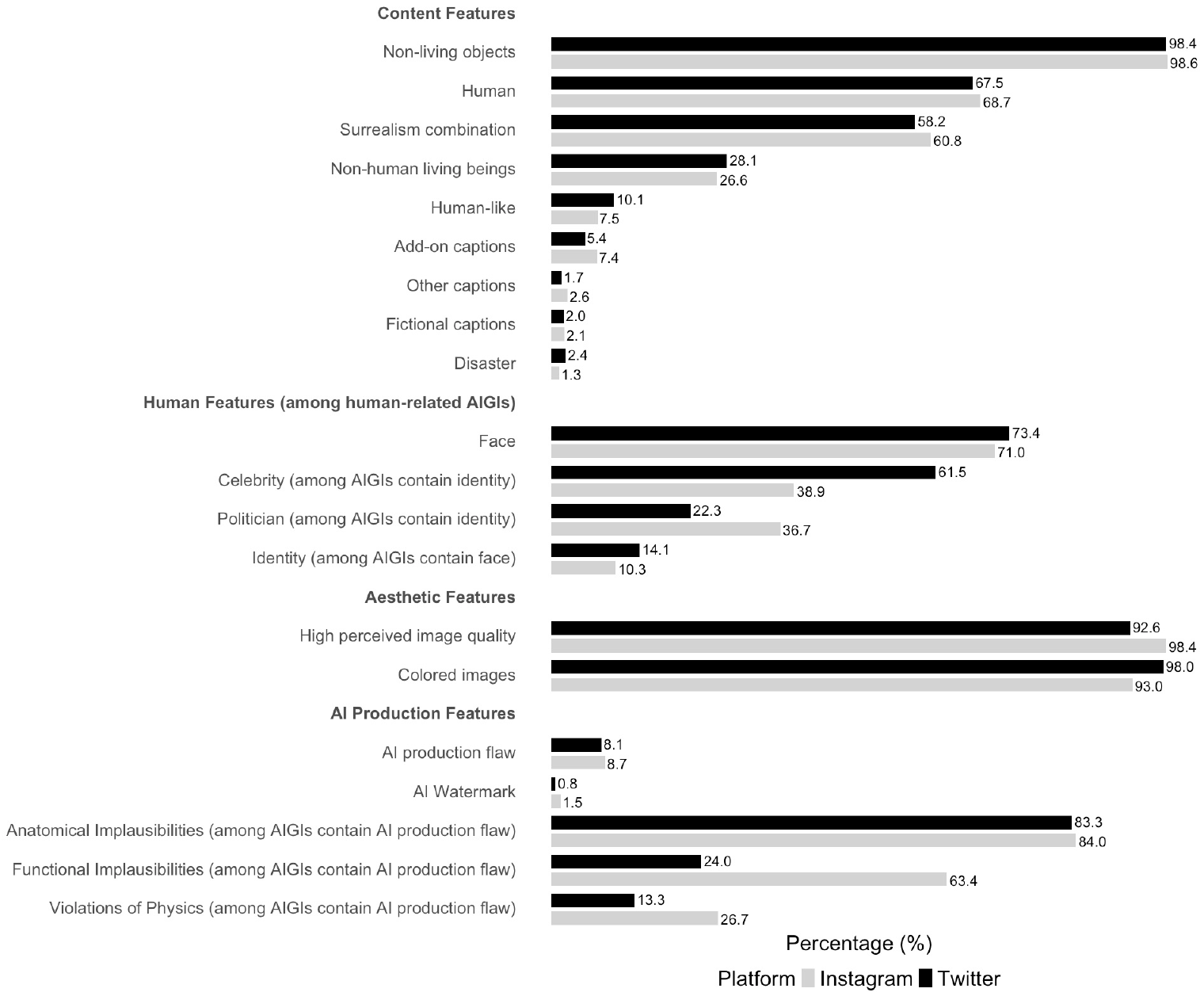}
  \caption{The Percentage of AIGI Visual Features}
  \Description{Percentages of AIGI Visual Features. The bar chart displays the distribution of various features in AIGIs on Instagram and Twitter, categorized by content, human-related features, aesthetic qualities, and production attributes. The chart includes comparisons across both platforms, with percentages representing the presence of each feature. Key features include content types such as non-living objects, humans, and surrealism combinations, as well as human-related features like faces and celebrity identities. Aesthetic features cover colored images and image quality, while production flaws and watermarks represent production characteristics. For some features, including face and identity-related features, their percentage is calculated based on the total number of their antecedent features.}
  \label{fig:vis}
\end{figure*}
\twocolumn

\onecolumn
\section{Comparison of features between Twitter and Instagram}\label{sec:Appendixd}
\begin{table*}[!ht]
    \centering
    \begin{tabular}{p{4cm}p{2.5cm}p{2.5cm}p{1cm}}
        \toprule
        \textbf{Variables} & \textbf{Twitter (\%)} & \textbf{IG (\%)} & \textbf{$\chi^2$} \\
        \midrule
        \textbf{Content features} & \textbf{N = 1,858} & \textbf{N = 1,785} & \\
        \midrule
        Human & 67.49 & 68.68 & 0.54 \\
        Human-like & 10.06 & 7.45 & 7.44** \\
        Non-human living beings & 28.15 & 26.55 & 1.08 \\
        Non-living objects & 98.44 & 98.60 & 0.07 \\
        Add on captions & 5.44 & 7.39 & 5.51* \\
        Fictional captions & 2.05 & 2.13 & 0.00 \\
        Other captions & 1.72 & 2.63 & 3.14 \\
        Disaster & 2.37 & 1.34 & 4.66* \\
        Surrealism combination & 58.23 & 60.78 & 2.35 \\
        \midrule
        \textbf{Human features} & \textbf{N = 1,254} & \textbf{N = 1,226} & \\
        \midrule
        Face & 73.37 & 70.96 & 1.66 \\
        \hspace{0.3cm} - Identity & 14.13  & 10.31 & 5.60* \\
        \hspace{0.7cm} - Politician & 22.31 & 36.67 & 4.73* \\
       \hspace{0.7cm} - Celebrity & 61.54 & 38.89 & 10.05** \\
        \midrule
        \textbf{Aesthetic features} & \textbf{N = 1,858} & \textbf{N = 1,785} & \\
        \midrule
        Colored images & 97.95 & 93.00 & 51.20*** \\
        High perceived image quality& 92.63 & 98.38 & 67.87*** \\
        \midrule
        \textbf{AI production features} & \textbf{N = 1,858} & \textbf{N = 1,785} & \\
        \midrule
        Production flaw & 8.13 & 8.68 & 0.30 \\
        Watermark & 0.75 & 1.51 & 4.06* \\
        \bottomrule
        \multicolumn{4}{p{12cm}}{\textit{Note: $p < .05^*$, $p < .01^{**}$, $p < .001^{***}$.}} \\
    \end{tabular}
    \caption{Comparison of features between Twitter and Instagram (N = 3,643)}
    \Description{Table 3 compares features of AIGIs between Twitter and Instagram, based on a sample size of 3,643 images (1,858 for Twitter and 1,785 for Instagram). The features are categorized into content features, human features, aesthetic features, and AI production features. Variables include the presence of humans or other entities, types of captions, surrealism, and human identities such as politicians or celebrities. The table also includes aesthetic metrics like color usage and image quality, as well as AI production features such as flaws and watermarks. Chi-square tests indicate mostly non-significant differences for some variables across the two platforms.}
    \label{tab:comparison_features}
\end{table*}
\twocolumn

\onecolumn
\section{Examples of AI Production Flaws}\label{sec:Appendixe}

\begin{figure}[h]
  \centering
  \includegraphics[width=0.7\linewidth]{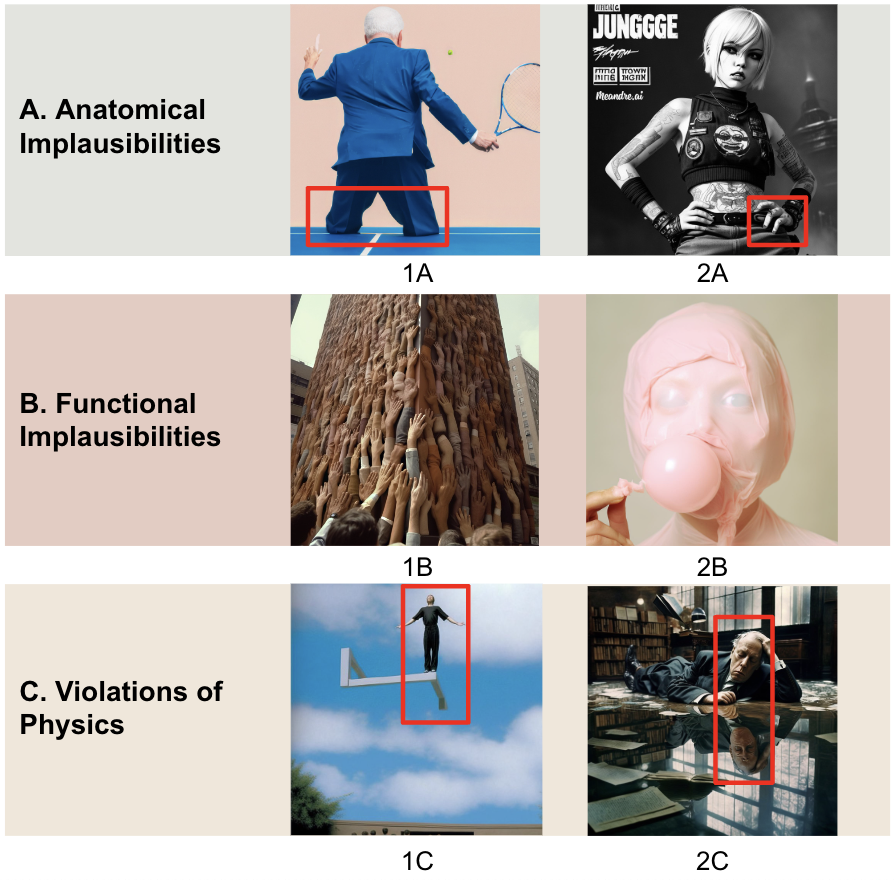}
  \caption{Examples of AI Production Flaw in AIGI}
  \label{fig:flaws}
  \Description{This figure contains examples of AIGIs with different types of production flaws. In the anatomical implausibilities category (Panel A), one example shows a man without legs, while another depicts a girl with only three fingers. In the functional implausibilities category (Panel B), one image features a tower constructed from piles of arms, and another shows a woman's face obscured by rubber-like materials. Finally, in the violations of physics category (Panel C), an example depicts a man without a shadow, while another highlights an incorrect reflection of the man on the floor.}
\end{figure}
\twocolumn

\end{document}